\documentclass[preprint]{aastex701}
\newcommand{\ltsim}{\raisebox{-1.0ex}{$\stackrel{\textstyle<}{\sim}$}}
\long\def\comment#1{}
\def\kms{km\,s$^{-1}$}

\def\goes{{\sl GOES}}

\def\hinode{{\sl Hinode}}

\def\p78{{\sl P78-1}}

\def\soho{{\sl SOHO}}

\def\sdo{{\sl SDO}}

\def\feix{Fe~{\sc ix}}
\def\fexii{Fe~{\sc xii}}
\def\fexiv{Fe~{\sc xiv}}

\def\fexviii{Fe~{\sc xviii}}
\def\fex{Fe~{\sc x}}

\def\civ{C~{\sc iv}}

\def\kms{km\,s$^{-1}$}

\def\etal{et~al.}

\begin{document}
%

\title{Surprisingly Large Doppler Shifts in \hinode\ EUV Imaging Spectrometer (EIS) Solar Spectra, Resulting from an Inconspicuous Small-scale
Jet in EUV Images}

\author[orcid=0000-0003-1281-897X]{Alphonse C. Sterling}
\affiliation{NASA/Marshall Space Flight Center, Huntsville, AL 35812, USA}
\email{alphonse.sterling@nasa.gov}

\author[orcid=0000-0001-9457-6200]{Louise K. Harra} 
\affiliation{Physikalisch Meteorologisches Observatorium Davos, World Radiation Center, 7260 Davos, Switzerland}
\affiliation{Institute for Particle Physics and Astrophysics, ETH Z{\"u}rich, 8092 Z{\"u}rich, Switzerland}
\email{Louise.Harra@pmodwrc.ch}

\author[orcid=0000-0001-7620-362X]{Navdeep K. Panesar} 
\affiliation{SETI Institute, 339 Bernardo Ave, Mountain View, CA 94043, USA}
\affiliation{Lockheed Martin Solar and Astrophysics Laboratory, 3251 Hanover Street, Building 252, Palo Alto, CA 94304, USA}
\email{npanesar@seti.org}

\author[orcid=0000-0002-5691-6152]{Ronald L. Moore} 
\affiliation{Center for Space Plasma and Aeronomic Research, \\
University of Alabama in Huntsville, Huntsville, AL 35805, USA}
\affiliation{NASA/Marshall Space Flight Center, Huntsville, AL 35812, USA}
\email{ronald.l.moore@nasa.gov}

\comment{

\author{\etal}

\author{Ronald L. Moore} 
\affiliation{NASA/Marshall Space Flight Center, Huntsville, AL 35812, USA}
\affiliation{Center for Space Plasma and Aeronomic Research, \\
University of Alabama in Huntsville, Huntsville, AL 35805, USA}

} 

\begin{abstract}

Strong EUV lineshifts in solar spectra are generally indicative of highly dynamic and explosive events that are easily detected 
in comparable-wavelength EUV images, with the strongest such line shifts (several 100\,\kms) occurring in solar flares.  Here 
we present observations of exceptionally strong lineshifts detected in \hinode/EUV Imaging Spectrometer (EIS) spectra outside
the time of a flare-like brightening, with 195\,\AA\ blueshifts of $\sim$200\,\kms\@. Although the likely culprit is 
too weak to register in \goes\ Soft X-ray fluxes, EIS pinpoints the source at the edge of an active region.  Solar 
Dynamics Observatory (\sdo)/Atmospheric Imaging Assembly (AIA) images and Helioseismic and Magnetic Imager (HMI) 
magnetograms show a nondescript small-scale eruptive event at this location.  We find this event likely to be an inconspicuous 
coronal jet, apparently triggered by converging/canceling magnetic flux patches, with plane-of-sky velocity 
$\sim$159$\pm 29$\,\kms.  AIA and HMI observations of this faint transient feature, together with observations of 
a slightly brighter jetting event near the same location an hour earlier, suggest that the strong EIS Doppler shifts are indeed due to 
a coronal jet that is hard to detect in AIA images.  These observations, together with other recent studies, show that  EUV 
Doppler maps are a much more sensitive tool for detecting small-scale eruptions than are EUV images, and those eruptions 
are frequently triggered by magnetic flux cancelation episodes.  Such-detected small-scale eruptions, that often produce 
small-scale coronal-jet-like features, might propagate into and help drive the solar wind.

\end{abstract}


\section{Introduction}
\label{sec-introduction}

Observations of the Sun in highly ionized iron lines, such as \fexii, are of emissions in the extreme ultraviolet (EUV) from 
plasma at temperatures characteristic of the corona.  On the solar disk, Doppler shifts in those lines blueward from their rest wavelengths are indicative of outflowing plasma from the corona.  Such observations of the solar corona are of high 
interest, since they can give insights into, for example, the driving of the solar wind, as well as into other fundamental solar 
processes. 

Solar flares typically show the largest blueshifts in EUV spectra, such as those obtained with the  \hinode/EUV Imaging
Spectrometer (EIS), reaching several 100\,\kms\ \citep[e.g.,][]{asai.et08,young.et13,doschek.et15}.  In non-flaring regions,
persistent or quasi-persistent outflows have been reported in EUV spectra.  For example, \citet{sakao.et07} observed what 
appeared to be outflows around the edges of some 
active regions in soft X-ray images from the X-ray Telescope (XRT) on \hinode, and then \citet{harra.et08} showed
that these apparent outflows had Doppler-detected blueshifts in EIS spectra of 20---50\,\kms, confirming that the apparent
outflows in the soft X-ray images were actually locations of material outflows.  Other such examples
are in, e.g., \citet{tian.et21} and in 
\citet{hinode.et19} (the chapter by D. H. Brooks).  Other than flaring plasmas themselves, a very large Doppler shift
(850\,\kms) has been observed in a flare spray that accompanied a flare \citep{young23}.  Also, a clear coronal jet observed
in soft X-rays by \hinode/XRT produced strong EIS-observed Doppler shifts of up to 240\,\kms\  \citep{morenoinsertis.et08}.
On the other hand, the source in the solar atmosphere of these line shifts can be a mystery 
when they are not obvious in coronal images, e.g.\ from the Atmospheric Imaging Assembly (AIA) on the Solar 
Dynamics Observatory (\sdo) satellite.
\citet{depontieu.et09} found EIS Doppler shifts (and also Doppler shifts from the \soho/SUMER instrument) that correlated 
well with observed chromospheric-jet/spicule velocities.
Their observations, however, did not allow for a specific one-to-one matching between the EIS observations and 
chomospheric-jet observations, due to incompatibilities between the spectrometers' spatial resolutions and cadences with the
size and dynamics of the chromospheric features, and therefore a definitive explanation for those spectral upflowing features is still wanting.  
Velocities of outflows of such features that lack confirmed identification can be less than 10\,\kms, and range up 
to $\sim$50---100\,\kms\ in some cases.

In addition, several observations with EIS have revealed the 
presence of outflowing regions in the corona that are localized (of extent $\ltsim$100$''$), and 
which are transient enough to persist less than the time for the implemented EIS raster scan, which can be a few tens of 
minutes in duration (as in our case here) depending on the observing program and the field of view.  
Despite having distinct EUV-Doppler signatures, the solar source of these transient outflows can also be difficult to identify 
definitively in AIA images.  This can be surprising, given that the EIS Doppler signatures themselves can be 
prominent against the background corona.  \citet{schwanitz.et21} examined 14 such localized and transient EIS outflows in quiet Sun and coronal holes that had 
velocities of some tens of \kms.  While they were able to identify characteristics of the source of the outflows in each 
case in AIA EUV images, doing so was a challenge due to the compactness and faint intensity of all of those features. We will
discuss these \citet{schwanitz.et21} events in \S\ref{sec-discussion}.  Earlier studies have similarly found EUV lineshifted features 
that are often difficult to interpret due to their small size and, for many such features, their generally weak signatures in EUV images 
\citep[e.g.,][]{madjarska.et09,innes.et13}.

Here we examine a feature with a particularly strong EUV line shift in EIS spectra, of $\sim$200\,\kms.  To our knowledge, it is the largest 
Doppler signature observed by EIS outside of flares, or obvious intensity features such as flare sprays or an obvious jet, 
and it is a mystery as to what non-strong-intensity solar activity can produce such large
Doppler shifts.  From the EIS spectral scan, we see that the event occurred near the edge of an active region where, as mentioned
above,  \citet{harra.et08} found outflow velocities of 20---50\,\kms\ in a more typical situation.  A casual 
inspection, however, of the AIA images, for example with the JHelioviewer software \citep{muller.et17}, at the time 
of the event does 
not show an obvious source for our high-velocity feature, as there are no large flares, and no obvious outstanding outflowing events, despite the strong
EIS signature.  Our objective is to identify in AIA images the source of this large EIS Doppler blueshift.

\section{Instrumentation and Data}
\label{sec-data}

EIS is an imaging spectrometer operating in the EUV range, with \citet{culhane.et07} providing full instrumental and operational 
details.  For this 
investigation EIS was scanning the target active region, NOAA AR\,12824, using a study named HPW021VEL260x512v2, 
consisting of several scans with each scan lasting about 60 minutes; the one with the primary event presented here ran from 
20:32:10 until 21:32:55 on 2021 May~21, and so lasted 60\,min~45\,s.  This covered several spectral 
lines, including the \fexii~195.120\,\AA\ line\@.  These spectra were taken with the $2''$ slit, 
and the exposure time was 20\,s.

Our imaging instrument for this investigation is \sdo/AIA, which observes the Sun in seven EUV wavelength bands with 
detectors of $0.''6$\,pixels and a maximum cadence of 12\,s, and also with two UV channels with a
cadence of 24\,s.  \citet{lemen.et12} provide full details of AIA, including temperature responses for each of the channels.  
We have inspected all of these channels for this study, and in the following we present results from selected representative 
channels, including in EUV: 171\,\AA, which is formed from \feix\ at a characteristic temperature of $\sim$0.6\,MK; 193\,\AA, from 
\fexii\ at 1.5\,MK; 211\,\AA, from \fexiv\ at 2.0\,MK; and 94\,\AA, which is largely from \fexviii\ at $\sim$6.3\,MK in flare observations, and
largely from \fex\ at $\sim$1.1\,MK for more quiet coronal observations \cite{odwyer.et10}.   We also present 
an image from
the UV 1600\,\AA\ channel, which shows upper photosphere continuum emission and flare emission in \civ\ near 0.1\,MK\@.   We 
supplement the AIA images with magnetic field data from \sdo's Helioseismic and Magnetic Imager 
\citep[HMI,][]{scherrer.et12}, with $0''.5$ pixels and which nominally takes a full-Sun line-of-sight magnetogram every 45\,s.

\begin{figure}
\hspace*{2.0cm}\includegraphics[angle=0,width=0.75\textwidth]{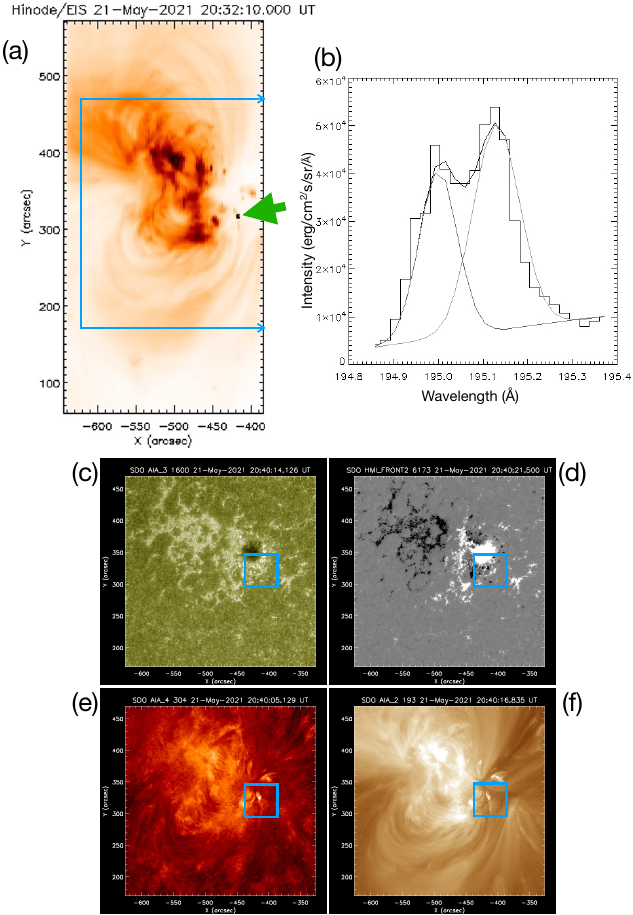}
\caption{
(a) \hinode/EIS raster image showing the observed active region, 
NOAA AR\,12824.  The raster started on the west edge of the field of view (FOV) at 20:32:10 UT 
on 2021 May~21, and finished at the east edge at 21:32:55\,UT, and the arrow 
points to the source of the strong Doppler-blueshifted spectral line shown in (b).  Blue lines show the FOV
of panels (c)--(e). (b) The EIS 195.12\,\AA\ spectrum from 20:39:51\,UT on the same day (histogram) with 
a two-Gaussian fit.  The peak on the left is strongly blueshifted compared to the peak on the right near the 
rest wavelength.  Panels (c---f) show images from \sdo/AIA and a magnetogram from
HMI of the region of the EIS strong-Doppler-shifted non-flare event, at the times given in the
tops of the respective panels: (c) AIA 1600\,\AA, (d) HMI line-of-sight magnetogram, (e) AIA 304\,\AA, 
and (f) AIA 193\,\AA\@.  The blue boxes show the FOV of Fig.\,\ref{big_blue_zoom2_zu}.
To within a minute, all four images in (d---f) are at the same time, and at the same time as the 
spectrum in (b). Solar images in all figures have north upwards and west to the right.  AIA and
HMI images in all figures are displayed rotated to 20:40\,UT on 2021 May 21.}
\label{bb_zu1}  
\end{figure}

The large EIS Doppler shifts were observed in AR\,12824 on 2021 May~21, where the spectra in 195\,\AA\ show the strong
shifts at two times during one of its scans across the active region, with the first one at 20:39:51\,UT and 
the second at 20:40:11\,UT\@.  Figure\,\ref{bb_zu1} shows the 
key EIS observations of the Doppler event, with panel\,(a) showing 
an image formed from one of the EIS raster scans at $195.12$\,\AA\ \fexii, formed at 1.5\,MK, where the scan started at 
20:32:10 UT\@.  The source of the strong Doppler shifts is rooted at or near the base of a large coronal
loop that spans the region, but located outside of the bright core loops of the active region; in Figure\,\ref{bb_zu1}(a) the green 
arrow points to the pixels (blackened in the image) in which the large Doppler shifts were observed.  EIS used its $2''$ slit for the raster, in which the slit scanned the 
pictured region from west to east, with about a one-hour duration for completion of the raster.

\clearpage

Figure\,\ref{bb_zu1}(b) shows a spectrum obtained 
when then the slit crossed the strong-Doppler-source region, producing a highly Doppler-shifted line at 20:39:51\,UT\@.  
Gaussians are fit to two components in this spectrum, one centered near the rest wavelength of 
195.12\,\AA\ \fexii\ line, and the other centered at 194.99\,\AA\@.  This line shift corresponds to a Doppler speed of 200\,\kms.

Full-Sun soft X-ray fluxes from the \goes\ spacecraft, which are used as proxies for solar flaring activity, show no emissions 
above the B2 level in the 1---8\,\AA\ channel between 20:10 and 21:30\,UT on 2021 May\,21 (shown in Fig.\,\ref{ronbun_zu3},
introduced below), and therefore there was no
activity that could be described as ``flaring" during the period of the observations near 20:40\,UT\@.  From the EIS data 
alone, it is not possible to ascertain the cause of the large Doppler shifts.  For further clues to the possible source of those
strong lineshifts, we turn to the AIA and HMI data.

\section{Results}
\label{sec-results}

\subsection{AIA Source of EIS-observed 20:40\,UT Event}
\label{subsec-eis}

Figure\,\ref{bb_zu1} panels\,(c)---(f) show an overview of AR\,12824 from AIA and HMI\@.  From the 
1600\,\AA\ image in (c) and the HMI magentogram in (d), there is a single positive-polarity sunspot in the region, 
with a mix of negative-polarity and positive-polarity flux clumps along the southwest side of that spot.  Comparing these panels with \ref{bb_zu1}(a) shows that the site of the 
strong EIS Doppler shifts is from within the location isolated by the blue box.

\begin{figure}
\hspace*{0.5cm}\includegraphics[angle=0,width=0.90\textwidth]{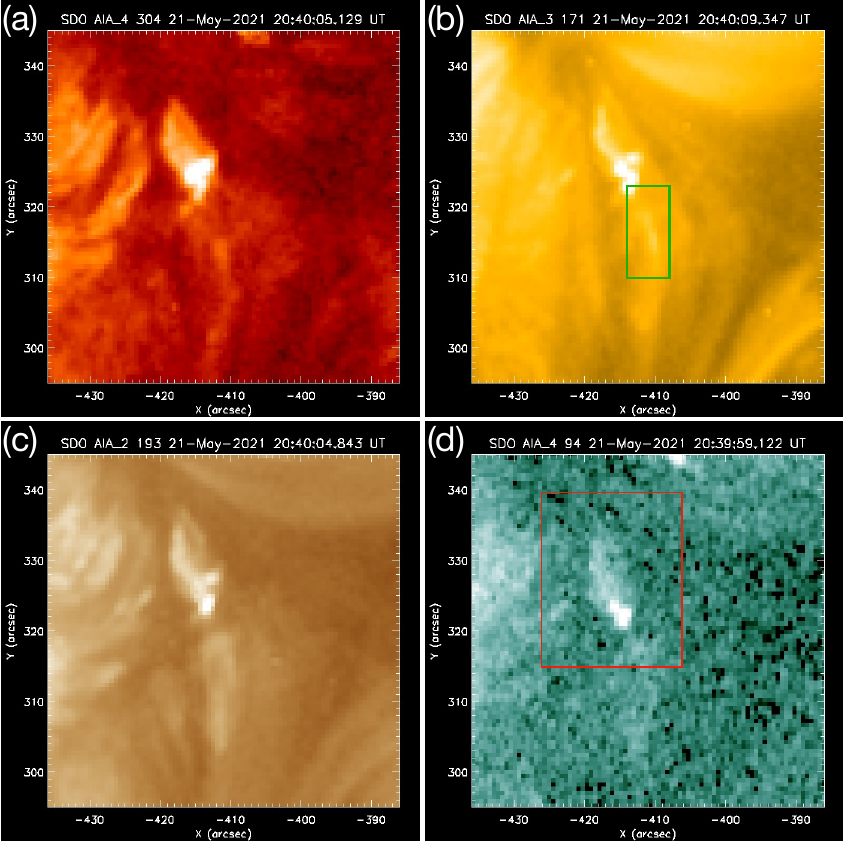}
\caption{
\sdo/AIA images of the zoomed-in region of the EIS strong-Doppler-shifted non-flare event, of the blue-boxed
FOV of Figs.\,\ref{bb_zu1}(c)---(f), at wavelengths: (a) 304\,\AA, (b) 171\,\AA, (c) 193\,\AA, and (d) 94\,\AA\@. To within 
a minute, these panels are at the same time as those in Figs.\,\ref{bb_zu1}(c)---(f).  In (b), the green rectangle shows the 
$y$-range covered in Fig.\,\ref{ronbun_zu3}(d) (the time-distance map in Fig.\,\ref{ronbun_zu3}(d) is produced by integrating
over the green box in the $x$ dirction).  In (d), the red box shows the FOV over which the AIA 94\,\AA\ lightcurve is 
calculated and plotted in Fig.\,\ref{ronbun_zu3}(b).  An animation accompanies the figure, running for 10\,s.}
\label{big_blue_zoom2_zu}  
\end{figure}

Figure\,\ref{big_blue_zoom2_zu} shows a zoom-in of the blue box's field of view (FOV) where the Doppler 
shifts originate, with the displayed images from the same time as those in the overview image of 
Figure\,\ref{bb_zu1}(c)---(f), but in the AIA 304, 171, 193, and 94\,\AA\ channels.   An accompanying 
animation shows a video sequence of the four panels over 20:00---21:00\,UT, and therefore covering the 
time of the strong Doppler shifts near 20:40\,UT\@.

\begin{figure}
\hspace*{2.0cm}\includegraphics[angle=0,width=0.7\textwidth]{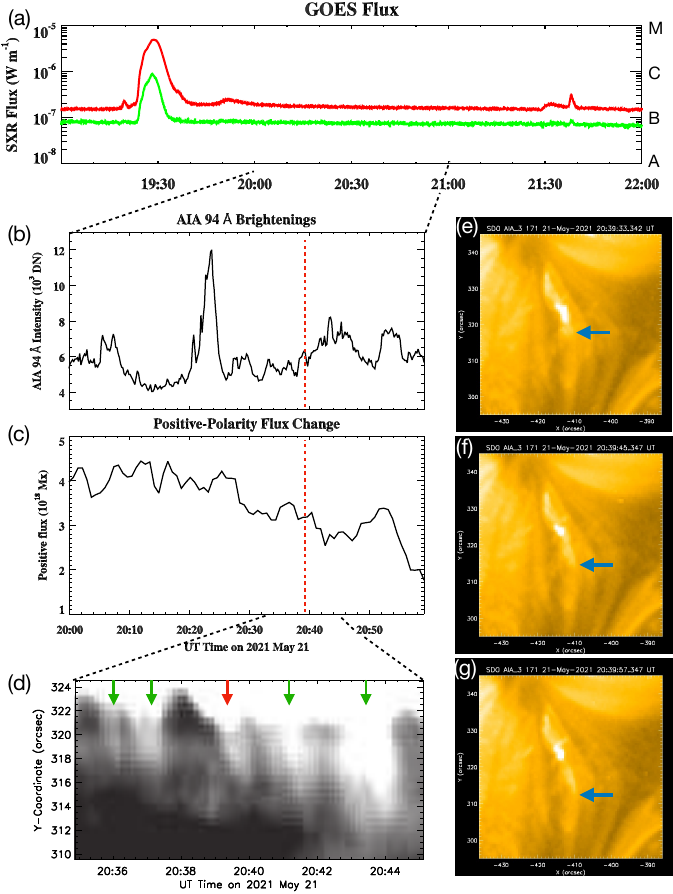}
\caption{
(a) \goes\ soft X-ray fluxes with time in the 1---8\,\AA\ (red) and the 0.5---4\,\AA\ (green) channels.  
(b) Variation of  AIA 94\,\AA\ intensity with time integrated over the red box of 
Fig.\,\ref{big_blue_zoom2_zu}(d), through the time of the AIA images in Fig.\,\ref{big_blue_zoom2_zu}.  
(c) Variation of positive magnetic flux with time in the positive-polarity flux patch in the yellow 
box of Fig.\,\ref{bb_hmi_zu}(g).  The red dashed line in (b) and (c) shows the time of the strong EIS Doppler shift.
 (d) A time-distance plot from AIA\,171\,\AA\ images, from the green 
 rectangle in Fig.\,\ref{big_blue_zoom2_zu}(b), 
 where the ordinate of that rectangle is the y-axis here, where we integrated over the width of 
 that green rectangle to increase the signal in this time-distance plot.  The resulting southward-directed flows
 in the Fig.\,\ref{big_blue_zoom2_zu}(b) green rectangle show up as bright tracks moving southward in time
 that are pointed out by the arrows; the red arrow is at the time of the strong Doppler feature of 
 Fig.\,\ref{bb_zu1}, and the green arrows show other south-flowing events.  Each of the arrowed 
 south-flowing events can be identified in the animation accompanying Fig.\,\ref{big_blue_zoom2_zu}, near
 the times indicated by the arrows. (e--g) A sequence of images from the Fig.\,\ref{big_blue_zoom2_zu}(b) 
 AIA 171\,\AA\ animation, showing a faint south-directed flow corresponding to the feature pointed to by the 
 red arrow in (d).  The green arrows track a location near the front of the moving feature, which is used in 
 estimating the flow speed given in the text.} 
\label{ronbun_zu3}  
\end{figure}

Figure\,\ref{ronbun_zu3}(a) shows \goes\ full-Sun soft X-ray lightcurves during the time of the event,
demonstrating that there is no indication of any flaring at the time of the large EIS Doppler shifts.  
Figure\,\ref{ronbun_zu3}(b) shows a lightcurve of the 94\,\AA\ flux variation with time over the region isolated 
by the red box in Figure\,\ref{big_blue_zoom2_zu}(d). From the Figure\,\ref{big_blue_zoom2_zu} video, small-scale 
brightenings do occur in that video's FOV at that red-box location, and the Figure\,\ref{ronbun_zu3}(b) 94\,\AA\ lightcurve
shows a comparatively strong brightening peaking at 20:24\,UT, and another brightening peak at about
20:44\,UT (with at least two sub-brightening peaks superposed on that broader 20:44\,UT peak).  These 
two brightenings, and several other brightenings over the displayed period of Figure\,\ref{ronbun_zu3}(b), are stronger than
the 94\,\AA\ intensity at the time of the strong Doppler shifts near 20:40\,UT, the AIA images for which 
time are displayed in Figures\,\ref{bb_zu1} and \ref{big_blue_zoom2_zu}.  Even though there are these few 
substantial brightenings in that Figure\,\ref{ronbun_zu3}(b) 94\,\AA\ lightcurve, those events are not seen in the 
Figure\,\ref{ronbun_zu3}(a) full-Sun \goes\ soft X-ray flux profile at the time of the large Doppler shifts, when the \goes\
background level is at about B2.

From the animation accompanying Figure\,\ref{big_blue_zoom2_zu}, there are no obvious expelled features that we would identify as 
``typical" EUV coronal jets from this region over the period 20:00---21:00\,UT\@.  (Here when we say that we see no ``typical EUV 
coronal jets," we mean that we would not have selected as typical coronal jets any of the small-scale dynamic features we see at these 
times in the AIA movies for our earlier detailed studies of EUV coronal jets, such as in the studies of 
\citeauthor{adams.et14}\,\citeyear{adams.et14}, \citeauthor{panesar.et16a}\,\citeyear{panesar.et16a}, \citeauthor{mcglasson.et19}\,\citeyear{mcglasson.et19}.)  We would expect any such jets to be expelled toward the south 
from the area that is bright in Figure\,\ref{big_blue_zoom2_zu} because coronal jets occurring at the edges of active regions follow 
field lines of the corona, which in this case would be spreading radially southward from the sunspot located to the north
of the brightening location \citep[e.g.,][]{sterling.et24a}, as indicated by the blue boxes in Figure\,\ref{bb_zu1}(c)---(f).  And 
indeed, more obvious southward-directed coronal jets do originate from this location at other times, e.g., at 19:20, 19:35, and 22:15\,UT,
that are outside of the time window of the Figure\,\ref{big_blue_zoom2_zu} animation.  (In \S\ref{subsec-earlier} we will show the 
19:35\,UT jet from this region.)  

There are, however, weak brightenings and outflows near the limit of detectability in AIA at the time 
of the strong EIS Doppler shifts near 20:40\,UT\@.  More specifically, in Figure\,\ref{big_blue_zoom2_zu} and its 
accompanying animation, there is a brightening in the 
center of the FOV of the figure, and the animation does show a series of intermittent transient outflows from that bright 
location moving southward over $\sim$20:33---20:44\,UT\@.  Figure\,\ref{ronbun_zu3}(b) shows the lightcurve in 
the AIA 94\,\AA\ channel integrated over the red box of Figure\,\ref{big_blue_zoom2_zu}(d), showing the brightenings at that location
as a function of time over about 20:35----20:45\,UT\@. 

We consider whether one of these intermittent transient brightenings might be the source of the EIS-observed blueshifts, as no other obvious candidate
presents itself in the Figure\,\ref{big_blue_zoom2_zu} animation.  To explore that possibility, we construct time-distance maps 
from the AIA images.  For the closest match to the EIS 195\,\AA\ spectral data, we would ideally
show an AIA 193\,\AA\ time-distance plot.  We created such a plot initially, but we found the tracks of the resulting southward-flowing 
features to be extremely feeble and hard to display.  In the 171\,\AA\ channel we found that these features,
although still weak, show enough contrast to render identifiable tracks in the time-distance plots,
and therefore this is what we choose to display in Figure\,\ref{ronbun_zu3}(d).  In constructing this plot, we express 
as distance on the ordinate axis the north-south extent shown by the green rectangle in Figure\,\ref{big_blue_zoom2_zu}(b), ranging 
from $y=310''$---$324''$.  But because the outflows are still weak in the 171\,\AA\ images, we integrate across the green box along 
the $x$ direction, over its $6''$ width spanning $-408''$~---~$-414''$.  This increases the intensity in the time-distance plots of the 
outflowing features, but sacrifices precision in the east-west location of the features.  With this, the outflows are detectable as 
bright southward (downward) pointing extensions in the Figure\,\ref{ronbun_zu3}(d) plot.     (For the 193\,\AA\ time-distance map 
we initially created, the tracks were still weak even with the intensity integrated in the $x$ direction.)

This Figure\,\ref{ronbun_zu3}(d) time-distance plot, covering ten minutes over 20:35---20:45\,UT, includes the period of interest 
for our study attempting to match the AIA observations with the strong spectral signature from 
20:39---20:40\,UT\@.  In the time-distance plots several features are visible.  Near the top of the panel is the track at $y = 324''$ in
Figure\,\ref{big_blue_zoom2_zu}(b), and that figure shows that the northeast corner of the green rectangle enters the bright point in that image; because
the intensity of the time-distance plot had to be increased enough to show the weak south-flowing features, and because that 
plot is displayed on a logarithmic intensity scale, that bright point is saturated for almost all times at that 
$y = 324''$ location in the time-distance plot.  But emanating from that bright point we can identify at various times several of 
the south-flowing features, with the more prominent ones indicated by the arrows at the top of the time-distance panel; the arrows 
are green, except for the one red one that points to the feature near the time of EIS strong Doppler blueshifts near 20:40\,UT\@.   Each of these
arrowed south-flowing features can be identified upon close inspection of the animation accompanying Figure\,\ref{big_blue_zoom2_zu}.
In the time-distance plot, the red-arrowed feature that corresponds to the strong EIS Doppler shifts is obviously weaker than the brightest 
south-flowing feature in that plot at 20:44\,UT, and even that 20:44\,UT feature is not particularly outstanding in the Figure\,\ref{big_blue_zoom2_zu}
animation.  This reflects that the strong-Doppler-producing event at near 20:40\,UT is faint and hard to detect in AIA 171\,\AA\ images.
Figure\,\ref{big_blue_zoom2_zu} also confirms that the feature at about 20:40\,UT is also faint in 193\,\AA\@.  That being said, the feature
is detectable in all four displayed channels; this includes the 94\,\AA\ channel, indicating that coronal-temerature plasmas are 
present in the feature.  It is unclear, however, whether the 94\,\AA-channel emission primarily comes from the \fexviii\ line formed 
at $T \sim 6.3$\,MK, which 
would mean the feature is similar in temperature to the 3---8\,MK temperature found for active region 
jets \citep{shimojo.et00,paraschiv.et22}; or from that channel's \fex\ line formed at 1.1\,MK \citep{odwyer.et10}, which would mean 
the feature is similar in temperature to the $\sim 1$---2\,MK found for coronal hole jets \citep{nistico.et11,pucci.et13,paraschiv.et15}.

We can check on whether the south-flowing feature near 20:40\,UT has a velocity signature that might correspond to that observed
in the EIS Doppler spectra, keeping in mind that the AIA images show the plane-of-sky velocity rather than the line-of-sight 
velocities from the Doppler shifts.

Determining the velocity of the outflows from the time-distance plots, however, is highly uncertain.  In Figure\,\ref{ronbun_zu3}(d)
the tracks for most of the southward-flowing features, including that of most interest near 20:39\,UT, are broad and have 
indistinctness of the edges, allowing for a variety of speeds to fit the trajectories.  This 
difficulty is in part due to our integration in the east-west direction across the green box in Figure\,\ref{big_blue_zoom2_zu}(c), 
and in part due to the faintness of the feature in the AIA 171\,\AA\ images that motivated that east-west integration.

We instead can estimate the speed more reliably directly from the Figure\,\ref{big_blue_zoom2_zu} animations.  
Figure\,\ref{ronbun_zu3}(e---g) shows a sequence from that animation in 171\,\AA, showing extension of about $6''$ 
over 24\,s, for a speed
of $\sim$182\,\kms.  We also made an estimate from 171\,\AA\ images over 20:39:21---20:39:45\,UT, finding a travel distance of 
about $5''$, which corresponds to 151\,\kms.   We also were able to do the same type of measurements
using 193\,\AA\ images, and found that we could track an element over 20:39:16---20:39:40\,UT that travels about 
$6''$, again giving 182\,\kms.  Doing the same in 193\,\AA\ images over 20:39:28---20:39:52\,UT indicates movement from 
about 322$''$ to 326$''$, yielding speed 121\,\kms. Taking the mean and standard deviation of these four values gives an average
observed plane-of-sky speed for the apparent outflows of $159\pm 29$\,\kms.  This is not far from the observed strong 
EIS Doppler velocity value of 200\,\kms.  Because the Doppler speed is the line-of-sight speed, which is almost normal to 
the solar surface, and that the estimates here from images are of the component of motion horizontal to the surface, the 
differences in speed estimates could could accounted for if the feature is inclined slightly more than 45$^\circ$ toward the 
vertical and toward Earth when observed by AIA and EIS\@.

\subsection{An Inconspicuous Coronal Jet at the Source of the Doppler Signature}
\label{subsec-pulsed}

We next address what these AIA brightenings and intermittent-transient-outflow features are, and what their cause may be. 
Typical coronal jets have a bright base, and a bright collimated spire that extends outward from that base region.
The features here appear similar to jets in that there is a bright region that could be a jet base, and apparent outflows
emanating from that base.  In the animation of  Figure\,\ref{big_blue_zoom2_zu} at 20:39:40\,UT in the 193\,\AA\
image (20:39:45\,UT in the 171\,\AA\ image), this southward extension from the base does appear to be a weak jet 
spire that is $\ltsim 10''$ in length, about $3''$ in width, and faint, particularly in 193\,\AA\@.  That being said, there is
also a southward extension in 94\,\AA\@.  In 304\,\AA, between about 20:39:05 and 20:42:29\,UT, there is nearly continuous
outflow that is generally toward the south, but not as collimated in the south direction as the feature in 
193 and 171\,\AA\@.  This appearance is similar to what was termed  {\it inconspicuous jets} in \citet{sterling.et22a};
those inconspicuous jets spanned lengths of $\sim$7---20$''$, and the event here is within that range.
Those inconspicuous jets had projected velocities of 19---46\,\kms; that is slower than the $\sim$159\,\kms\ projected
velocity found for the current event, although this difference could be because the current event occurred at the edge
of an active region, and so was formed in stronger magnetic field than the inconspicuous jets of \citet{sterling.et22a},
which occurred in coronal holes.

Another characteristic of coronal jets is that they frequently occur with their bases at the sites of converging 
and canceling opposite-polarity magnetic flux elements \citep[e.g.,][]{adams.et14,panesar.et16a,panesar.et18a,mcglasson.et19,muglach21}.  Several studies \citep[e.g.,][]{panesar.et16a,panesar.et18a,sterling.et16b,sterling.et17,mcglasson.et19} argue that many, if not most or all, coronal 
jets form when magnetic flux cancelation builds up and triggers the eruption of a small-scale magnetic flux rope, often 
carrying cool material in the form of a minifilament erupting along with the field \citep{sterling.et15}.  The magnetic field of the 
erupting miniflament/flux rope undergoes 
interchange reconnection (which \citeauthor{sterling.et15}~\citeyear{sterling.et15} call ``external reconnection" in the jet-producing
situation) with surrounding far-reaching coronal magnetic field, to make the
jet spire that is propelled outward along the reconnected far-reaching field. 

\begin{figure}
\hspace*{0.5cm}\includegraphics[angle=0,width=0.90\textwidth]{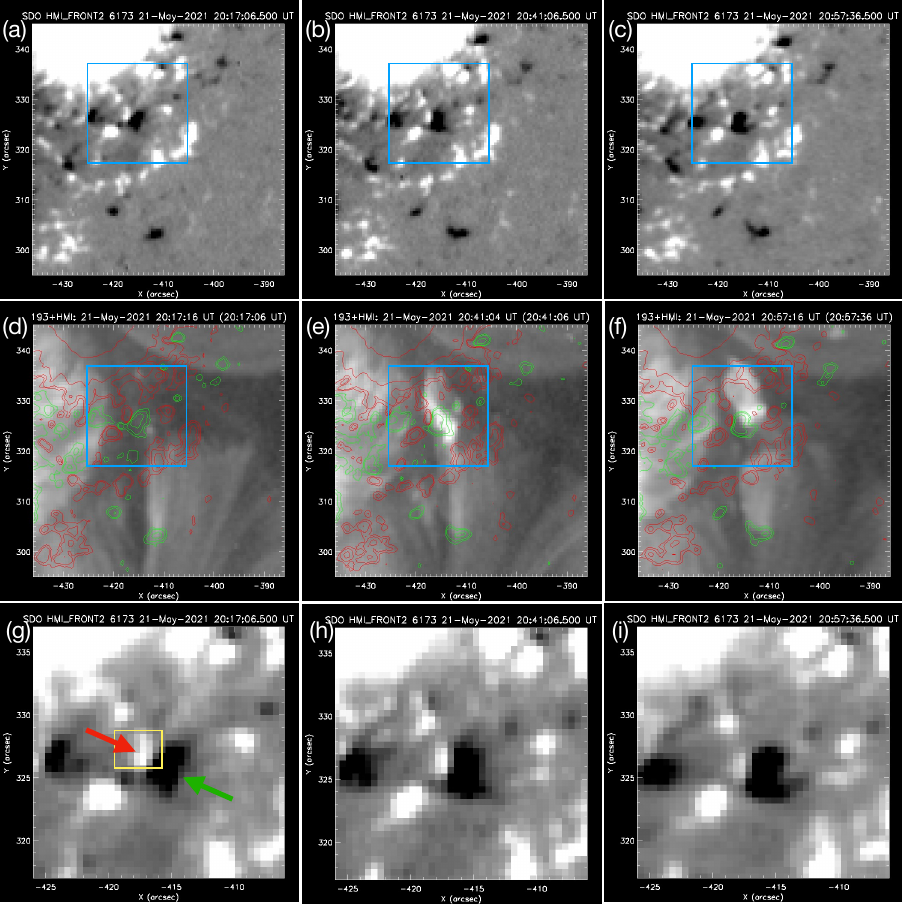}
\caption{
HMI magnetograms of the strong-Dopper-shift region.  Panels\,(a)---(c) show the magnetic evolution of the region with the
same FOV as in Fig.\,\ref{big_blue_zoom2_zu}, at three different times: 20:17, 20:41, and 20:58\,UT on 2021 May 21,
with white and black denoting positive and negative polarities, respectively.  Panels\,(d)---(f) show AIA 193\,\AA\ images at 
the same times, with contours of the magnetograms overlaid, where the image
times are at the top of the panels without parentheses and the magnetogram times are at the tops of the panels in parentheses.
Red and green contours denote positive and negative polarities, respectively, and the contour levels are 
at plus and minus 30, 100, 300, and 1000\,G\@.  Panels\,(g)---(i) show the magnetograms of (a)---(c), with the FOV of the
blue boxes of panels\,(a)---(f).  Panel\,(g) shows a positive-polarity magnetic-flux patch (red arrow) that progressively 
becomes canceled with a negative-polarity magnetic-flux patch (green arrow) with time over the course of the displayed time period.
It is plausible that this flux cancelation triggers the dynamics resulting in the strong Doppler shifts of Fig.\,\ref{bb_zu1}.  
An animation accompanies the figure, showing the FOV of panels (a)---(c), covering the time span of 19---21\,UT on 2021 May 21, 
and running for 6\,s.} 
\label{bb_hmi_zu}  
\end{figure}

Inspection of the videos of Figure\,\ref{big_blue_zoom2_zu}, along with the time-distance plots of 
Figure\,\ref{ronbun_zu3}, suggest that the transient outflows toward the south originate with the brightening apparent in
the boxed region of Figures\,\ref{bb_zu1}(c)---(f), and that brightening is also visible near the center of the four panels 
of Figure\,\ref{big_blue_zoom2_zu}.  We examined HMI magnetograms of the region, as presented in 
Figure\,\ref{bb_hmi_zu}. Figures\,\ref{bb_hmi_zu}(a)---(c) show
magnetograms of the FOV of Figure\,\ref{big_blue_zoom2_zu}.  Comparing with concurrent 193\,\AA\ images
(Figs.\,\ref{bb_hmi_zu}(d)---(f)) shows that the brightening occurs where there are mixed-polarity flux patches.  
Arrows in Figure\,\ref{bb_hmi_zu}(g) point to two such patches that undergo flux cancelation before, during, and after the
blueshift event, with the 
positive-flux patch progressively consumed by the negative-flux patch over Figures\,\ref{bb_hmi_zu}(g---i).   
Figure\,\ref{ronbun_zu3}(e) plots the positive-polarity patch's flux with time (as bounded by the yellow box in 
Fig.\,\ref{bb_hmi_zu} (g)), showing that it indeed does decrease over 20:00---21:00\,UT, consistent with it undergoing 
flux cancelation.

As pointed out above, the faint intensity extension toward the south at about 20:39\,UT could be a weak jet spire. 
And indeed this seems to be an inconspicuous jet, given that the magnetic setup is similar to that of many jets.  Therefore, we 
expect that magnetic flux cancelation results in magnetic reconnection that makes one or more small-scale magnetic flux 
ropes that each erupts and forms a faint EUV jet.  This process could occur repeatedly and somewhat
intermittently, manifesting as a series of small-scale jetting events at the times of each of the arrows in 
Figure\,\ref{ronbun_zu3}(d).

If the events are indeed faint EUV jets, then they might be too small for easy detection of an erupting minifilament, assuming that
cool material exists on the small erupting flux ropes.  In our observations we can detect several small-scale candidates for being 
such erupting minifilaments over the time period of Figure\,\ref{ronbun_zu3}(d), that appear sometimes as absorbing features and
sometimes in emission moving away 
from the base brightening in the 304\,\AA\ videos in the animation accompanying Figure\,\ref{big_blue_zoom2_zu}, 
e.g.\ over $\sim$20:27---20:43\,UT\@.   Additionally, 
there may be cases where eruptions occur but without a cool-material minifilament being clearly visible, as occurs in some 
large-flare-producing eruptions, and which also seems to occur in some jet-producing eruptions \citep{kumar.et19}.  The features we 
observe here, however, are small and faint compared to jets of many of our other studies, and it may be that easy-to-discern erupting
minifilaments are too small to be detected in the AIA images.  This is similar to the situation with some faint X-ray jets \citep{sterling.et22a}, 
and with the smaller jet-like features referred to as ``jetlets" \citep[e.g.,][]{raouafi.et14,panesar.et18b}.

\subsection{19:35\,UT Event}
\label{subsec-earlier}

We have inspected this same region at earlier times to see whether similar jet-like features might be apparent that could 
provide further insight into the 20:40\,UT inconspicuous jet that we suspect produced the large EIS Doppler shifts.  We 
found that at about 19:35\,UT, i.e.\ roughly an hour before the EIS-observed event, there is a similar episode from 
nearly the same photospheric footpoint location.  EIS data are not available at this location at the time, and so we 
only examine the AIA and HMI data.  

\begin{figure}
\hspace*{0.0cm}\includegraphics[angle=0,width=1.0\textwidth]{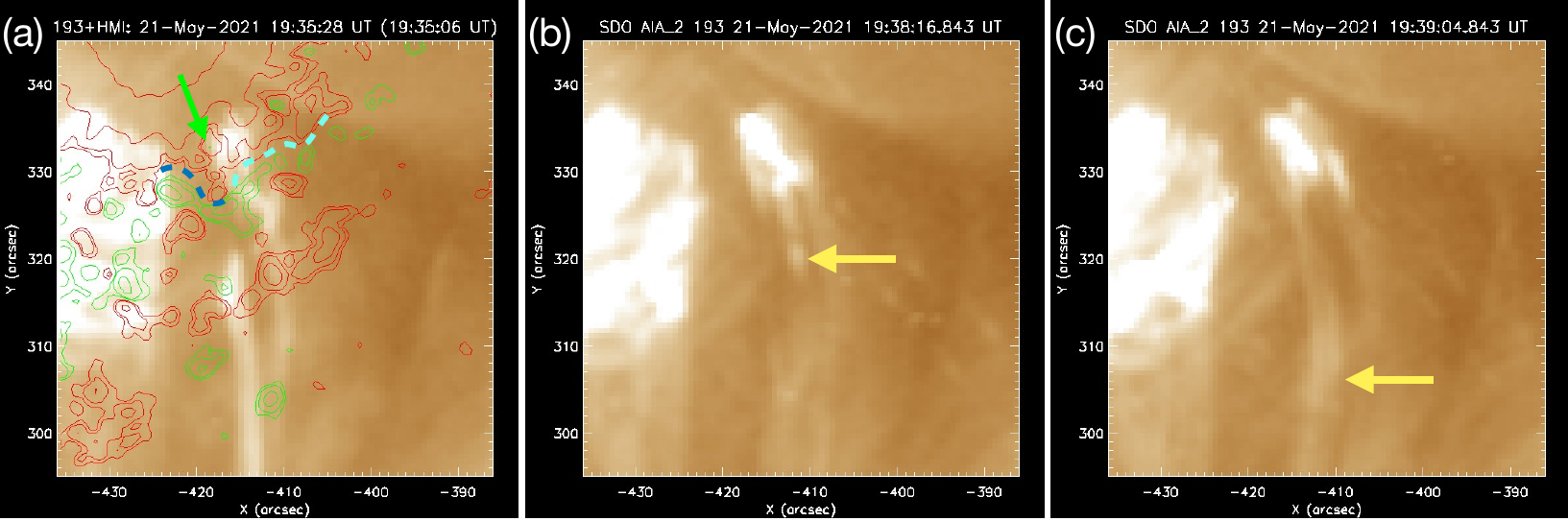}
\caption{
AIA\,193\,\AA\ images of the same location on the Sun as for the 20:40\,UT event in previous figures, but showing an event near 
19:35\,UT that was stronger in AIA images than that 20:40\,UT event.  In (a), HMI magnetogram contours as those in 
Fig.\,\ref{bb_hmi_zu}  are overlayed onto the 193\,\AA\ image.  This 19:35\,UT event much more obviously appears as a coronal jet, 
and emanates from about the same location as the 20:40\,UT event.  This jet extending outward toward the south, in the same direction as 
the weaker extensions from the 20:40\,UT event shown in the green box of Fig.\,\ref{big_blue_zoom2_zu}.  This 19:35\,UT 
coronal jet-like feature originates near a
mixed-polarity location, near the neutral lines indicated by the dashed line segments in (a) and/or from the location
where a weak negative polarity element is growing at the location of the arrow in (a).  An intensity clump of the jet tracked by
the arrows in (b) and (c) extends outward with a velocity of $\sim$200\,\kms.  This 19:35\,UT event may be a stronger version of
the 20:40\,UT event, where in that 20:40\,UT event EIS was able to discern the strong blueshifts of Fig.\,\ref{bb_zu1}(b) 
even though any outflowing jet-like material from that 20:40\,UT event was near the detection capabilities of 
AIA in the images and accompanying video of Fig.\,\ref{big_blue_zoom2_zu}.  An animation accompanies this figure, 
running for 2\,s.}
\label{big_blue_19ut_zoom2_zu}  
\end{figure}

Figure\,\ref{big_blue_19ut_zoom2_zu} shows three frames from AIA 193\,\AA\@.  Morphologically, this event is similar
to our EIS-observed event an hour later, but it is brighter in the AIA 193\,\AA\ images.  This 19:35\,UT 
event is more obviously a coronal jet that is expelled outward and to the south; this is particularly 
apparent in the video accompanying the figure.   This 19:35\,UT jet appears to be rotating, specifically between 
19:36 and 19:41\,UT; if viewed from the south looking northward along the spire, it appears as if this rotation would be
in the counterclockwise direction.  Such rotation (spin) of the spire is commonly seen in coronal jets \citep[e.g.,][]{pike.et98,moore.et15,panesar.et22,kayshap.et24}.
This fits in with the minifilament-eruption idea, if the pre-eruption minifilament flux rope has twist that it releases through 
the external reconnection \citep[e.g.,][]{shibata.et86,moore.et15,panesar.et16a,sterling.et24}.  Numerical modeling supports 
that this proposed mechanism is viable \citep{li.et25}.

Beyond the morphological similarities with the 20:40\,UT event, we can follow 
a clump of material in the AIA images of this 19:35\,UT event pointed to by the 
arrows in Figure\,\ref{big_blue_19ut_zoom2_zu}(b) and (c), and over the 48\,s between those frames the outward speed is 
$\sim$197\,\kms; we measured the motion of such clumps over three intervals between 19:38:16 and 19:39:04\,UT, and 
found for the speed a mean and 
standard deviation of $219\pm 40$\,\kms.   These values are very close to the speeds we estimated from AIA images in the 
EIS-observed event an hour later, and all of these are similar to the magnitude of the EIS Doppler shift in 
Figure\,\ref{bb_zu1}(b) of $\sim$200\,\kms. This similarity supports that the two events are similar phenomena, but with the 
19:35\,UT one brighter in AIA images than the 20:40\,UT one.  

Figure\,\ref{big_blue_19ut_zoom2_zu}(a) overlays an HMI magnetogram (with contours as in Fig.\,\ref{bb_hmi_zu}(d---f)) onto an AIA\,193\,\AA\ image of the 19:35\,UT event, 
with neutral lines traced by the dashed lines.  A positive-polarity element pointed to by the green arrow in the frame is 
dynamically changing, and apparently undergoing flux cancelation, in the accompanying animation in the HMI movie accompanying 
Figure\,\ref{bb_hmi_zu}; that animation spans 19--21\,UT, and thus includes this 19:35\,UT event in addition to the 20:40\,UT event.  These magnetograms
thus show that there are mixed polarities likely undergoing flux cancelation in the base region from which the 19:35\,UT coronal 
jet emanates, similar to the situation with the 20:40\,UT EIS-observed event.

Although stronger in AIA intensity than the 20:40\,UT EIS-observed event, this 19:35\,UT feature is still comparatively 
weak in intensity compared to those used in typical coronal jet studies, and it is spatially too small to establish fully its origin.  
Nonetheless, based on these AIA images and HMI magnetograms, 
this 19:35\,UT jet is consistent with having an origin as the jets describe in \citet{sterling.et15} and \citet{panesar.et16a}, 
discussed earlier.  This lends support to our suggestion that the 20:40\,UT event that creates the strong EIS Doppler shifts is similarly 
created in the same physical manner as a typical (brigher) EUV coronal jet, but too faint to see well in AIA 193\,\AA\ intensity images.

Returning to the time-distance plot in Figure\,\ref{ronbun_zu3}(d) of the time period around the 20:40\,UT event, this shows
repeated transient-outflow episodes, making multiple bright sloped trajectories; therefore, if these are all inconspicuous coronal jets, 
then they are repeating jets.   Some active region jets \citep{sterling.et16b,sterling.et17} show similar repeated-jet phenomena; for 
example, Figure\,8 of  \citet{sterling.et17}  shows numerous coronal jets originating at the
same canceling neutral line over about a two-hour period ($\sim$14:00---16:00\,UT on 2015 January 14 in their Fig.\,8), due to eruptions
of a series of minifilament ``strands."  In that case, the jetting continues as long as the flux cancelation occurs.  A similar phenomenon 
occurs with non-active-region jets, forming homologous jets \citep{panesar.et17}, but much farther spaced in time than the active 
region jets, presumably because of the weaker magnetic fluxes in quiet Sun and coronal holes compared to active regions.

\section{Discussion}
\label{sec-discussion}

We have proposed that the strong, 200\,\kms\ Doppler blueshifts observed by EIS (Fig.\,\ref{bb_zu1}(b)) on 2021 May 21 
near 20:40\,UT result from an inconspicuous coronal jet.  This jet is faint in AIA images, and therefore is difficult to confirm
as a jet without close inspection.  Upon such close inspection, the overall morphology of the bright jet base, and
outflows appearing as a spire (albeit a faint one) are consistent with coronal jet morphology.  Moreover, the base region is in
a magnetically mixed-polarity location that is undergoing flux cancelation, similar to many coronal jets.  These factors support
that we are observing a difficult-to-detect EUV coronal jet, one that is occurring at the time the EIS spectrometer slit fell on this
location.  In addition, our measured plane-of-sky speed of that jet of $159\pm 29$\,\kms\ is comparable to the 200\,\kms\ 
Doppler outflow (line-of-sight) velocity measured by EIS, lending support to the suggestion that this inconspicuous coronal jet made 
the EIS strong Doppler shifts.  The 20:40\,UT feature that we investigated is one of a series of transient outflows from the
same base region, as evidenced by the arrows in Figure\,\ref{ronbun_zu3}(d).  We suspect that all of these transient outflows
are similar coronal jets.

Although an erupting minifilament, if present here, is too small to be clearly seen in the AIA images, overall the observations 
are consistent with these events resulting from minifilament/flux rope eruptions, where the minifilament/flux ropes would be 
built up and triggered to erupt by magnetic flux cancelation between a comparatively large negative-polarity magnetic flux 
patch and a smaller 
positive-polarity magnetic flux patch (Fig.\,\ref{bb_hmi_zu}). This is similar to the magnetic evolution that occurs at the 
base of brighter coronal jets, but in this case the jet spire is faint in the AIA images.   A brighter jetting event occurs 
from the same location about an hour earlier, at $\sim$19:35\,UT, but without EIS observations.  In several ways, that 19:35\,UT 
event is morphologically similar to the AIA transient outflow that accompanies the EIS blueshifts, and has similar plane-of-sky 
speed as the EIS-observed 20:40\,UT event; both are $\sim$150---200\,\kms\ in AIA images, and this is close to the 
$\sim$200\,\kms\ line-of-sight Doppler velocity measured by EIS\@. The 19:35\,UT event is brighter in AIA 193\,\AA\ intensity 
images than the EIS-observed 20:40\,UT event, and has more of the morphological appearance of a coronal jet.  This 
provides further support that the EIS-observed event is produce by a coronal jet, but with a spire that is faint in EUV 
intensity.  This assumes the observed jet-like outflows 
have a line-of-sight component comparable to what we see in the plane-of-sky AIA images.  In addition, the HMI 
magnetograms show flux cancelation occurring at the base of both the 19:35\,UT and the 20:40\,UT events, along with 
AIA brightenings at magnetic neutral lines at the base of the event; these features are consistent with those seen 
in many coronal jets.

The outstanding finding of this study is that a small scale coronal jet, one that is subtle and difficult to discern when looking at 
only AIA images and videos, can produce remarkably strong signatures in EUV Doppler spectral data.  Despite the EIS Doppler 
signatures of Figure\,\ref{bb_zu1} being among the largest reported by EIS outside of flares, the feature that we identify as the 
best candidate for producing those Doppler shifts is a faint, inconspicuous coronal jet in AIA, the magnetic base of which is
consistent with the jet being driven by a minifilament eruption that was built and triggered to erupt by magnetic flux cancelation, but 
that was not substantial enough to make a more obvious coronal EUV jet.  Similar features in the AIA images result is a series of
transient outflows (arrows in Fig.\,\ref{ronbun_zu3}(d)), plausibly driven by similar minifilament eruptions (or a series of erupting minifilament
strands).

Similar cases of strong Doppler shifts with weak EUV-imaging signatures have been found in several earlier studies.  
\citet{young15} observed a number of strong Doppler-shifted localized features in a coronal hole with EIS 195\,\AA\ spectral scans, which
he termed ``dark jets" because they did not show an obvious corresponding outward-moving feature in AIA 193\,\AA\ images.  Those
dark jets had Doppler velocities ranging from 57---107\,\kms, and appeared as bright points in the images, but did not appear as obvious coronal
jets.  \citet{schwanitz.et21} similarly observed 14 outflow events in EUV Doppler data, seven of which had at most weak signatures 
in corresponding imaging data.  \citet{sterling.et22a} examined in greater detail five of the \citet{schwanitz.et21} events, and found 
all of them to be weak coronal jets whose evolution was consistent with the minifilament-eruption mechanism, even though they were 
inconspicuous in the EUV images; having been confirmed as coronal jets, those were the jets that \citet{sterling.et22a} termed as 
``inconspicuous jets."  \citet{harra.et23} found substantial EIS Doppler outflows originating near a sunspot umbra, for which 
there was no clearly obvious source in EUV images from AIA or from the EUVI imager on {\it Solar Orbiter}.

In light of the findings here however, that a particularly strong EIS-detected coronal transient outflow can be produced 
by a coronal jet that is weak in intensity in AIA images, there is the outstanding question of whether the so-far unidentified 
sources of the \citet{young15} dark jets and/or
the Doppler feature of \citet{harra.et23} might be made by even fainter coronal jets that are beyond detection in AIA images.

Most studies directed specifically at coronal jets \citep[e.g.,][]{moore.et13,sterling.et15,panesar.et16a,mcglasson.et19} deliberately 
pick out for study easy-to-identify coronal jets with obvious spires in imaging data.  None of the \citet{young15} dark jets, or the inconspicuous jets of 
\citet{schwanitz.et21} and \citet{sterling.et22a}, would have been identified in investigations of coronal jets in such studies due
to their weakness in the imaging data, yet all of these events were nonetheless conspicuous in the EIS Doppler data.  That 
is the same for the subtle transient-outflow EUV feature that we propose as the source of the large EIS Doppler shifts here.  Therefore, it is apparent 
that EUV Doppler imaging is a powerful device for detecting robust outflow events that are difficult to detect in AIA EUV images.  Also however, 
the strong Doppler feature of \citet{harra.et23} was not identified as a coronal jet even upon close inspection, and so at least some such 
line shifts may be made by a non-coronal-jet mechanism.

Coronal jets and jet-like events are very commonly observed in EUV \citep[e.g.,][]{raouafi.et16,sterling.et24}, occurring at rates 
of at least some hundreds per day.   In contrast, the detection in EIS Doppler images of dark jets or inconspicuous jets is
relatively infrequent \citep[][]{young15,schwanitz.et21,sterling.et22a}.  This is because EIS obtains Doppler data from scans 
of its narrow slit over a relatively limited FOV\@.  Because of this, even though coronal jets are comparatively frequent, EIS 
has to have its narrow slit crossing the correct location at the right time to detect them.  As an example, the brightenings 
and dynamic outflows flows less than five minutes after the our 20:40\,UT Doppler observations, so at about 20:43---20:44\,UT in
Figure\,\ref{ronbun_zu3}(c), were much much more substantial in AIA in the video accompanying 
Figure\,\ref{big_blue_19ut_zoom2_zu}.  Similarly, the event at 19:35\,UT also was much more substantial in AIA images than 
the EIS-observed 20:40\,UT event.  Presumably these stronger-AIA-intensity events would have made Doppler signatures, 
and perhaps with line shifts comparable to or greater than that shown in Figure\,\ref{bb_zu1}(b), similar to the 240\,\kms\ found 
for the earlier-mentioned jet of \citet{morenoinsertis.et08},  but the EIS slit was in 
scanning mode and so not at the same location at the times of the larger brightenings.    Imaging spectrometers with substantial 
FOV and much higher cadence, such as the upcoming {\sl MUSE} mission, and also the EUVST instrument on the upcoming 
{\sl Solar-C} mission, potentially will show many more such outflows.
Such observations potentially will be vitally important for understanding the source of the solar wind, which may be powered by jet-like
phenomena, as recently discussed \citep{raouafi.et23,sterling.et24,chitta.et25}.

\begin{acknowledgments}

We thank the referee for providing comments that improved the paper.  ACS and RLM received funding from the 
Heliophysics Division of NASA's Science Mission Directorate 
through the Heliophysics Supporting Research (HSR) Program, and ACS also was supported by the 
MSFC \hinode\ Project.  LKH thanks the Swiss National Science Foundation (SNSF) for the support of 
grant 200021\_219368.  NKP acknowledges support from NASA’s SDO/AIA (NNG04EA00C) grant, NASA's  
HCSI (80NSSC25K7028) grant, and NASA’s HSR (80NSSC24K0258) grant.

\end{acknowledgments}

\bibliography{ms}{}

\clearpage


\end{document}